# Bell's Theorem, Many Worlds and Backwards-Time Physics: Not Just a Matter of Interpretation

Paul J. Werbos
*Program Director for Quantum, Molecular and High Performance Modeling and Simulation
Engineering Directorate, National Science Foundation[1]
Arlington, Virginia, US 22230*

**Abstract.** The classic "Bell's Theorem" of Clauser, Holt, Shimony and Horne tells us that we must give up at least one of: (1) objective reality (aka "hidden variables"); (2) locality; or (3) time-forwards macroscopic statistics (aka "causality"). The orthodox Copenhagen version of physics gives up the first. The many-worlds theory of Everett and Wheeler gives up the second. The backwards-time theory of physics (BTP) gives up the third. Contrary to conventional wisdom, empirical evidence strongly favors Everett-Wheeler over orthodox Copenhagen. BTP allows two major variations – a many-worlds version and a neoclassical version based on Partial Differential Equations (PDE), in the spirit of Einstein. Section 2 of this paper discusses the origins of quantum measurement according to BTP, focusing on the issue of how we represent condensed matter objects like polarizers in a model "Bell's Theorem" experiment. The backwards time telegraph (BTT) is not ruled out in BTP, but is highly speculative for now, as will be discussed.

## 1. EVERETT VERSUS ORTHODOX COPENHAGEN

Conventional wisdom has drifted into the comfortable *assumption* that Bell's Theorem[1-3] is basically just about philosophy, not about physics. In many ways it simplifies life just to assume that all the various formulations of physics are equivalent to each other for all practical purposes. This section provides a relatively simple counterexample, mainly as a way of calling for us to re-examine our assumptions more carefully.

<u>**1a. Quick Review of the Two Theories**</u>

This subsection will give a definition of what the two theories are. In effect, I am defining terminology here for the expert reader. For a nice introductory explanation of what the two theories are about, *Scientific American* ran an interesting story on Hugh Everett in December 2007, with references for the general reader.

In classical physics, from the time of Newton to the time of Einstein, the main goal of physics was to find a complete model of objective reality, a model of how the universe works, a model from which we can deduce detailed predictions for measurements and for the outcome of experiments. In flat space, before gravity is accounted for, Einstein proposed that the universe is governed by PDE which could be represented simply as:

$$\dot{\varphi}_i = f_i(\underline{\varphi}, \underline{\nabla \varphi}, \underline{\nabla^2 \varphi}), \qquad i=1,N \tag{1}$$

where N is the number of classical field components out there in objective reality. I will refer to this as the *PDE theory of physics*. Please note that the PDE theory of physics is very different from the Lorentzian version of classical physics, which combines continuous fields with perfect point particles; Lorentzian physics is the limit of quantum mechanics as $\hbar \to 0$, but Einsteinian physics is different. Equation 1 is capable of representing second-order dynamical systems, and so on, but it is not necessary to get into those details in this section.

---

[1] The views herein are not anyone's official views, but this does constitute work produced on government time.

Most physicists even today are initially taught to believe a competing theory which I will refer to as the *orthodox Copenhagen* theory (with apology to some of the friends of Heisenberg like Duerr and Stapp who are anything but orthodox). The main tenets of orthodox Copenhagen are:

(1) *Unreality.* Physics, in this view, is not about describing how reality works, but about making predictions. We are warned that the mathematical objects used in quantum theory – like the wave function, $\Psi$ -- represent *our knowledge* of the state of the universe. We are warned that it is impossible and meaningless to try to write equations describing the *actual objective* state of the universe. For example, Heisenberg's "principle of uncertainty" is not really a principle of uncertainty; it is a principle of *indeterminism*. The point is *not* that we can not measure the position and momentum of a particle at the same time; rather, Heisenberg claimed that particles do not actually *possess* a precise position and momentum at the same time in reality.

(2) *A three-step recipe for predicting any experiment.* Since all of the physics lies in how we make predictions, the recipe for prediction is the logical core which defines any theory. Modern physics has actually evolved a rather huge and not entirely unified collection of recipes, but the orthodox recipe[4-6] consists of just three parts: (a) encode your knowledge of the initial conditions of the experiment into a wave function $\Psi(t_-)$, where $t_-$ is the initial time; (b) translate your initial knowledge, $\Psi(t_-)$, into knowledge $\Psi(t_+)$ about the outcome or equilibrium of the experiment, by using the "Schrödinger equation":

$$\dot{\Psi} = iH\Psi \tag{2}$$

and (c) predicting the expected value of any measurable outcome variable *m* by calculating:

$$<m> = \Psi^* M \Psi \tag{3},$$

where M is the measurement operator corresponding to the specific variable being measured.

(3). *The wave function as a vector in Fock-Hilbert space.* The recipe is only meaningful, of course, when we specify what $\Psi$ is, and describe the recipes for how to encode initial knowledge into $\Psi$. $\Psi$ at any time t is essentially just the *set* of complex-valued functions:

$$\Psi_N(\underline{x}_1, s_1, \underline{x}_2, s_2, ..., \underline{x}_N, s_N) \tag{4}$$

for all values of N from zero to infinity. Each argument $\underline{x}_k$ is a vector in three-dimensional space, which we think of crudely as a possible position of particle number k. Each argument $s_k$ represents a collection of discrete variables, such as the spin or identity, describing particle number k. Quantum theory requires that $\Psi$ has a square norm of one, when $\Psi$ is considered as a vector in Fock-Hilbert space. $\Psi^*$ is just the Hermitian conjugate of that vector.

In his PhD thesis under John Wheeler in the 1950's (reprinted by DeWitt[7]), Hugh Everett of Princeton proposed a breath-taking alternative to the orthodox Copenhagen theory. He proposed that we should return to the notion of objective reality. He proposed a theory of physics in which the wave function $\Psi(t)$ actually expresses the state of objective reality at time t. Equation 2 is the actual dynamical law governing the universe. In his thesis, he proved a theorem which, he claimed, showed how equation 3 could be *derived* as a consequence of this dynamics. According to this theory, the true universe is actually infinite dimensional, as implied by equation 4; thus the three dimensional universe or "world" that we see in everyday life is only one of the many "worlds" which exist side by side. This was the original "many-worlds" theory of physics.

**1b. Empirical Comparison of the Two Theories**

In the 1970's, most of us believed that the difference between these two theories was mainly philosophical. For example, DeWitt[7] argued that they would make different predictions for certain kinds of measurement experiments – but only if we could use the entire volume of the known universe or more to set up the apparatus.

Only in the 1990's did I realize that we had missed something very fundamental here. For myself, the realization came from watching what was happening in the exciting new field of quantum computing[8,9].

The original idea for true quantum computing came from a series of technical papers by David Deutch of Oxford, inspired by the many-world theory of physics. Crudely speaking, the idea was that we could harness the reality of parallel universes in order to do massively parallel computing. Prior to Deutch, Richard Feynman and others had made comments like "there is room at the bottom" and "maybe we could use quantum mechanics somehow in computing," but Deutch was the one who actually suggested a way to do it. (Failure to give proper credit to people like Deutch is a serious pathology in today's culture, one which seriously interferes with understanding of the substantive issues.)

In the first wave of enthusiasm for quantum computing, people discussed how the superposition of wave functions representing different "worlds" could be used for a kind of parallel computation. When proposals first came to NSF, the initial reviews were highly divided. Many favorable reviews came from leading theoretical physicists, ironically because the orthodox Copenhagen theory clearly predicted that this wild idea would actually work. But the condensed matter physicists disagreed.

The initial proposals for quantum computing simply would not have worked, because they did not understand the kind of "linearity" imposed by statistical many-body effects in the real world. People doing practical, empirical prediction of electronics and quantum optics in the real world learned many years ago that the $\Psi$ by itself simply cannot describe our knowledge adequately. The orthodox Copenhagen recipe does not work. Instead, most practical people today rely on an object called the density matrix $\rho$, which may be roughly but adequately defined via the functional integral:

$$\rho(t) = \int \Psi \Psi^* \Pr(\Psi, t) \, d^\infty \Psi \tag{5}$$

Notice that this definition relies heavily on $\Pr(\Psi,t)$, the probability that the true wave function of the universe at time t is actually equal to a particular possible value $\Psi$. It is only meaningful if we implicitly accept Everett's theory that $\Psi$ represents a possible state of objective reality. Just as $\Psi$ is a vector in Fock-Hilbert space, $\rho$ is a matrix over that same space, as suggested by the simple matrix multiplication in equation 5.

This implicit assumption pervades the calculations which actually work in the practical trenches of empirical reality. As an example, Mandel and Wolf[10], the most standard text in quantum optics today, states on page 480: "… let us consider the density operator $\rho^\wedge$ of the field. This is the natural object for describing the state of a system that is not in a pure quantum state, but is describable by an ensemble of quantum states $|\psi_i\rangle$, with probabilities $p_i$." Likewise, the classic text by Kadanoff and Baym[11], which underlies modern calculations in quantum electronics[12], relies heavily on the assumption that the state of *our knowledge* about the universe can only be described by a density matrix. The grand canonical ensemble (the foundation of modern quantum statistical mechanics) relies on statistical reasoning about possible states $\Psi$, which implicitly assumes the many-world theory and equation 5.

Success in quantum computing became possible only when a group of cross-disciplinary people understood this challenge, and developed a new way to formulate quantum computing based on "pseudo-pure states[9]" of the density matrix $\rho$. The first working quantum computers were based on nuclear magnetic resonance designs by Gershenfeld et al, based on this concept.

**In summary, the many-worlds theory actually works, in leading to predictions which fit with the massive empirical data of modern electronics and quantum optics. The orthodox Copenhagen theory does not. It is possible to "repair" the orthodox Copenhagen theory, simply by replacing all references to $\Psi$ by references to $\rho$. (I hereby define that modified theory as $\rho$-Copenhagen.) But this leads to an obvious question: if the dynamics of the complex object $\rho$ can be *derived* from the many-worlds theory (as in quantum statistical mechanics), why should we really believe it is the true foundation of what we observe in the physical universe? Why not simply return to reality here? And why do people continue to use textbooks at any level which promulgate the orthodox Copenhagen view?**

## 1c. Quick comments on other quantum theories

All across the fields of high-energy physics, theoretical physics, quantum foundations and axiomatic quantum field theory, many alternative formulations of quantum field theory have been proposed through the years. It would require more than one book to do justice to them all.

Still, in order to fit classic Bell's Theorem experiments[3], they must give up one of the three classic axioms discussed in the work of Clauser et al. All of the usual theories which I am aware of give up either reality or locality (or both); thus they are cousins of orthodox Copenhagen or of many-worlds, or perhaps even equivalent to them.

Many years ago, Louis de Broglie[13] and David Bohm tried to formulate an Einsteinian theory of physics, based on an object called the "Q potential." In discussions with DeBroglie circa 1970, I pointed out that this only fits empirical data if the Q potential itself is defined as a full-fledged vector over Fock space, just like $\Psi$. This ends up reinventing the many-world theory through the back door, by way of Q. Modern work on the Bohm-Hiley approach formulates it as a nonlocal, many-worlds class of theories.

It is disturbing that Everett's theory, orthodox Copenhagen, and $\rho$-Copenhagen all treat time as a privileged coordinate. This grossly violates the spirit of special relativity. Streater and Wightman[14] proposed an alternative formulation of physics, in the spirit of Copenhagen, which is transparently relativistic. However, many of us have troubles with assumptions such as the requirement that the particle/fields must have a finite $L^2$ norm integrated over all space and all time; it is natural (and traditional) to assume that particles are more or less localizable in space, but to assume that across time presents certain difficulties. The program of axiomatic physics based on Streater and Wightman has concentrated so far on attempting to provide a more well-defined version of the orthodox Copenhagen theory of physics. That program has not yet gotten far enough to replicate quantum electrodynamics (QED).

A complete review would certainly say more about the Wick transform formulation[15] and the functional integral and path integral formulations of quantum field theory; however, for now, it is better that I simply refer to the work of Weinberg[6] and others, who have discussed the close relation between those formulations and the canonical orthodox approaches.

## 2. QUANTUM MEASUREMENT, BELL THEOREMS, BACKWARDS-TIME PHYSICS AND CONDENSED MATTER REPRESENTATIONS

### 2a. What's Wrong With Everett's Theory and How to Fix It

Everett's theory of physics[7] includes two major elements: (1) the claim that the dynamics of the universe are completely specified by equation 2 (after H is specified); and (2) the claim that equation 3 can be deduced from equation 2, without introducing any new physical assumptions. The second claim is the real problem here.

Everett's theorem has certain analogies to Aquinas's "proofs" of the existence of God, as well as to certain arguments in classical thermodynamics about the reasons behind the arrow of time.

With regards to the first analogy – it is fully understandable why many of us want to believe that the laws of measurement should be deduced from the underlying dynamics of the universe. The reasons for wanting to believe that this can be done are as compelling to us as Aquinas's goals were compelling to him. But when we are strongly motivated to believe the conclusion, we need to be very careful to ask how much we believe the proof as such. Can we think of something like a simple counterexample to the proof as such (without needing to decide whether we believe what the proof is getting at)?

This leads to my second analogy. The simplest forms of classical thermodynamics[16] start out by assuming a set of N equations:

$$\frac{dn_i}{dt} = c_i n_1^{k_{i1}} n_2^{k_{i2}} ... n_N^{k_{iN}} \qquad (6)$$

which are derived directly from assuming *time-symmetric* laws of collision between the underlying N species of elementary particle or molecule. From this system, they immediately deduce how entropy proceeds forwards in time. In other words, we have started from a set of assumptions about the universe which were totally symmetric with respect to time. From those assumptions, we have deduced a description of macroscopic statistics which is totally asymmetric with respect to time. How could that be possible, without inserting additional hidden assumptions somehow, either in classical thermodynamics or in Everett's case? (Equation 2 is perfectly symmetric with respect to time, T, except for some unexplained "twelfth decimal point" superweak interactions.) Because equations 2 and 6 are symmetric with respect to time, why can't we deduce the exact same conclusions as classical thermodynamics and Everett do, *for the backwards direction of time?*

For classical thermodynamics, I would claim that the explanation is well-known. We know that the classical laws of collision describe a dynamical system which is allowed to *start out* in any state (e.g. any nonnegative values of the numbers $n_i$), but tends towards a stable fixed point which depends on the initial values of conserved quantities like energy. The problem lies in explaining how the universe could have *started out* in a state which is so far away (it appears) from that stable fixed point. This is explained by invoking *specific boundary conditions* – the initial state of the universe – *in addition to equation 6*.

Following this analogy, I would claim that we can only derive the true equations of quantum measurement by also exploiting what we know about the boundary conditions of the universe. We should modify Everett's theory by postulating that all aspects of physics at finite time – microscopic or macroscopic, measurement or other – are governed by time-symmetric statistics; only the infinite-time boundary conditions impose any asymmetry. Instead of assuming that measurement must always obey equation 3 precisely, we should work to derive a higher-order understanding which allows for the possibility that quantum measurement may not always work exactly as traditionally advertised. We should work from a truly time-symmetric understanding of many-worlds physics, which does not just *assume* time-forwards statistics for all macroscopic phenomena. This approach I will define as the **many-worlds version of backwards-time physics (MW/BTP).** Of course, BTP does not assume that everything runs back wards in time only! It merely allows for effects which are not purely forwards-in-time only, as in the mathematics of mixed forwards-backwards stochastic differential equations[20].

## 2b. Two Examples Where MW/BTP Changes Things: Big Crunch and Polarizers

The cosmology of the Big Bang and the Big Crunch provides a very straightforward example of the huge difference between MW/BTP and the original Everett theory. Huw Price[17], a well-known philosopher of physics, pointed out years ago that Hawkings' *first edition* of his "Brief History of Time," proposed a cosmology in which the Big Crunch at the end of time would be a perfect mirror image of the Big Bang. This was the cleanest and most economical way to incorporate what was known from cosmology and astronomical data, and the general assumptions of astronomers, at that time. Equation 2 certainly fits the spirit of that original theory.

Yet, as Price tells it, Hawkings made huge changes to the book in a later edition, in order to reflect the frantic anxieties of those who realized how it would radically conflict with today's quantum physics. It would not conflict with quantum dynamics (equation 2), but it would conflict with the orthodox version of quantum measurement. According to Price, this change added lots and lots of epicycles to the theory, but it was felt necessary to put forward a theory consistent with the current understanding of quantum theory.

But what happens if we shift to MW/BTP? In that case, the boundary conditions at future time have just as much status as the boundary conditions at past time. We would *predict* that the practical "laws" of quantum measurement would in fact shift sign in regions of the universe where the boundary conditions having the greatest direct effect are different. This prediction may or may not be true, in the end, just as Hawking's theory about a Big Crunch may or may not be true. According to MW/BTP, we really could *test* the original version of Hawking's theory by building apparatus capable of looking for "signals" which come to us from the far future. Price's writings do not propose a highly efficient way to do those tests; for that, we need to translate the concepts of BTP into engineering tools suitable for modeling and designing a more efficient apparatus. That leads us into our second example – how to represent simple objects like a Polaroid polarizer[18], for use in cosmological experiments or in the analysis of Bell's Theorem experiments?

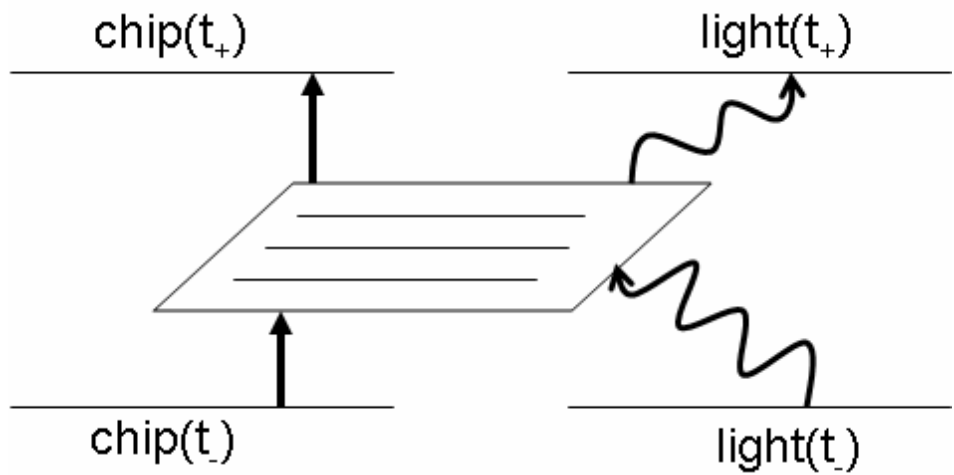

Figure 1. Picture of What a Simple Polaroid Polarizer Does

Figure 1 illustrates the challenge we face here. For simplicity, consider the case where the incoming and outgoing light consists of one or zero photons, linearly polarized, and where other sources of incoming or outgoing radiation do not have a significant effect. For practical purposes, the polarizer is basically just a set of narrow wires embedded in a matrix which allows some dissipation of heat and electricity. I use the expression "light($t_-$)" to refer to a simple combination of variables – presence or absence of photon, and the polarization of the photon. I use the expression "chip($t_-$)" to refer to the far more complex ensemble of variables needed to specify the state of the entire polarizer at time $t_-$.

In the orthodox analysis of this simple system, we would use condensed matter physics in order to estimate:

$$\Pr(light(t_+) | light(t_-)) = \int \Pr(light(t_+) | light(t_-), chip(t_-)) \Pr(chip(t_-)) d^N chip(t_-)$$
(7)

We would assume that the initial state of the polarizer, Pr(chip($t_-$)), is uncorrelated with any of the other inputs to the experiment or any of the other components; it is given by the usual Boltzmann distribution. The various thermodynamic or optical Green's functions would be used to provide the machinery for calculating this term. The validity of this assumption is essentially the same as the specific form of "causality" assumed in the main "Bell's Theorem" experiments[2,3].

Nevertheless, we simply cannot expect equation 7 to remain valid in the general case described by Huw Price, for example. From the viewpoint of MW/BPT, equation 7 is simply *assuming* that we cannot have macroscopic correlations from future to past. More precisely – according to MW/BPT, the arrow of time comes from boundary conditions in the larger world. The boundary condition which leads us to expect time-forwards results in Figure 1 is the boundary condition which provides *free energy* at time $t_-$, and allows us to emit a photon of visible light at time $t_-$. But the polarizer itself does not "know" about this boundary condition. In this simple situation, the polarizer itself is a completely time-symmetric object. If light emitted from a distant future galaxy were to send a photon of visible light back to this same polarizer, we would have every reason to expect the same behavior *backwards in time* that we normally see in the forwards direction.

How can we model the polarizer as a time-symmetric object, to do full justice to MW/BPT? If our goal were simply to compute the usual condition probability correctly, in the more general case:

$$\Pr(light(t_+) | light(t_-)) =$$
$$\int \Pr(light(t_+) | light(t_-), chip(t_-)) \Pr(chip(t_-)) d^N chip(t_-) *$$
(8)
$$\int \Pr(light(t_+) | light(t_-), chip(t_+)) \Pr(chip(t_+)) d^N chip(t_+) / Z$$

where Z is the usual partition function in combining Pr(chip($t_-$)) and Pr(chip($t_+$)). In other words, we can apply the usual approaches of Markhov Random Fields (MRF)[19], by accounting for the fact that we are facing uncorrelated random inputs/properties in both boundaries here, the future time boundary and the past time boundary.

The mathematics of MRF has been well developed, in a rigorous and general way, in engineering and in computer science. It is basically just the more general case of lattice statistics, a subject which is well-developed in physics[15,20,21]. Here I am proposing that we think of the entire space-time continuum as a kind of continuous lattice, and calculate the equilibrium statistics which result when we account for all the boundary conditions (future and past) in a balanced way. This is the kind of calculation required in the analysis of mixed forwards-backwards dynamical systems[22].

Concretely, however, when we analyze something like a Bell's Theorem experiment, we usually want to use some kind of lumped parameter model. A Bell's Theorem experiment is essentially a kind of circuit, and we can use a finite number of variables to describe the "input/output" behavior of each major element of the circuit, like the polarizers. In orthodox or classical physics, we *assume* that the circuit is a perfect feedforward system running from past to future; in that case, the simple conditional probability $\Pr(light(t_+)|light(t_-))$ is all we need to know about the polarizer, for a simple experiment like the usual Bell's Theorem experiment. But in BTP, we are trying to calculate the equilibrium statistics of a network connected in both directions. We must characterize the circuit elements by calculating clique probabilities, which may be thought of as the joint probability of the inputs we are looking at conditional upon their interaction within the circuit element and integrated over the uncertainties we are not really looking at. In essence, we want to calculate

$$"\Pr(light(t_+), light(t_-))" =$$
$$\int \Pr(light(t_+), light(t_-) | chip(t_-)) \Pr(chip(t_-)) d^N chip(t_-) * \quad (9)$$
$$\int \Pr(light(t_+), light(t_-) | chip(t_+)) \Pr(chip(t_+)) d^N chip(t_+) / Z$$

This provides the more complex information we need at the "circuit level" in order to predict the outcomes of the entire experiment

In theoretical classical and orthodox physics, if there is an incoming photon of linear polarization $\theta_1$, and the polarizer is aligned at angle $\theta_p$, there are only two allowed outcomes in the lumped parameter model:
1. No photon comes out at time $t_+$, with conditional probability $\sin^2(\theta_1-\theta_p)$;
2. A photon comes out at time $t_+$ with polarization $\theta_p$, with conditional probability $\cos^2(\theta_1-\theta_p)$

No photon of visible light is allowed at all at time $t_+$ if there is none coming in at time $t_-$. Small adjustments are introduced in experimental work, to allow for imperfections of the polarizer.

In the MRF approach, the lumped parameter description of the polarizer would allow for at least three significant possibilities:
1. A photon of polarization $\theta_p$ at both times, $t_-$ and $t_+$, with a clique probability of $c_0$.
2. A photon of polarization $\theta$ at just one of these times, with a clique probability of $c_1*\sin^2(\theta-\theta_p)$ in either case

3. A photon of polarization θ at one of the times $t_\pm$ and a photon of polarization $θ_p$, with a clique probability of $c_2*\cos^2(θ-θ_p)$ in either case

Again, this kind of model would not predict weird behavior in ordinary classical types of experiments, because of the effect of the boundary conditions – such as the lumped parameter model of the usual sources of light. But with certain special types of experiments, like the Bell's Theorem experiments[3], a unique source of light is used which does lead to the emergence of weird behavior.

## 2c. PDE/BTP: Another Alternative Theory

In this paper, I have argued that the MW/BTP theory of physics is a natural evolution of physics, firmly grounded in the empirical reality of circuits and quantum optics. It is also an easy evolution in a way, because we still reply on the familiar "Schrödinger" equation (equation 2) and the familiar Hamiltonian operators. In this section, I will briefly mention a far more radical alternative.

It starts with three basic questions. Once we get so far as MW/BTP, do we really need all those infinite dimensions of space? Is there any way to get back to a theory which lives up to the full spirit of special relativity, in which time is just one of four dimensions? Can we even find a way to get back to the path proposed by Einstein, using simple partial differential equations (PDE) to describe how the universe works inside of the boundary conditions provided by cosmology? These are the questions I actually started from here, beginning in 1973 when I was still in graduate school[23] and was the first to propose the backwards-time interpretation of quantum mechanics.

The original "Bell's Theorem" of Clauser et al[2] does allow for this possibility, once we give up equation 7. But it is a very complicated business to try to rederive the whole of quantum field theory, with nuclear forces and all, starting from PDE and equation 9. Since 1973, I have basically been working backwards myself (when time permitted), and trying to chart out the path which leads forward from today's formulations to this more radical long-term possibility. As in this paper, there is a natural path of evolution which *starts out* by addressing unresolved problems within today's physics, and thereby gets us closer to a return to reality[24].

For this paper, I will focus on the specific question of Bell's Theorem, which does not require discussion of nuclear forces or even of how to treat fermions. It is enough to consider electromagnetism[25], which we have discussed at great length with Yanhua Shih, one of the lead experimenters in this field and in optics-based quantum information technology.

According to PDE/BTP (and also according to Willis Lamb), the photon does not exist. Electromagnetism follows Maxwell's Laws (except of course for its sources and sinks). A garden-variety "photon" is simply a solution of Maxwell's Laws between two essential boundary conditions – the source which "emits" it and the sink which "absorbs" it. The photoelectric effect is merely the mirror image in time of the time-forwards emission of light; in BTP, there is no paradox at all in the empirical fact that both boundary conditions are important. The quantization of light is due to properties of the source and the sink, which are far more complicated[24] than electromagnetism as such.

Of course, Maxwell's Laws allow for the possibility of solutions which are not just superpositions of solutions between individual sources and sinks. This is what gives rise to coherence phenomena. Glauber and others have proven exact quantum-classical equivalences (see chapter 11 of Mandel and Wolf[10]) which provide a rigorous basis for understanding coherence effects. There are special cases, in orthodox time-forwards analysis, when a Glauber-style analysis seems to lead to negative probabilities[26]; however, I view those cases as special experiments, like Bell's Theorem experiments, where future-time boundary conditions need to be accounted for in order to arrive at a complete and consistent picture of what is really going on.

It is occasionally said that a perfectly coherent state of light still has a non-zero "variance," calculated with quantum-mechanical field operators. However, if the term "variance" is defined as in statistics (with the equivalent field operators), this is not correct. It is important to subtract out the mean.

## 2d. What Happens in a Bell's Theorem Experiment According to BTP

*Both versions* of Backwards Time Physics (BTP) would rely on equation 9 in analyzing what happens in a Bell's Theorem experiment. In this section I will run through the analysis of what happens here, according to BTP in general – but it is easier to think in terms of PDE/BTP, which provides a very concrete picture.

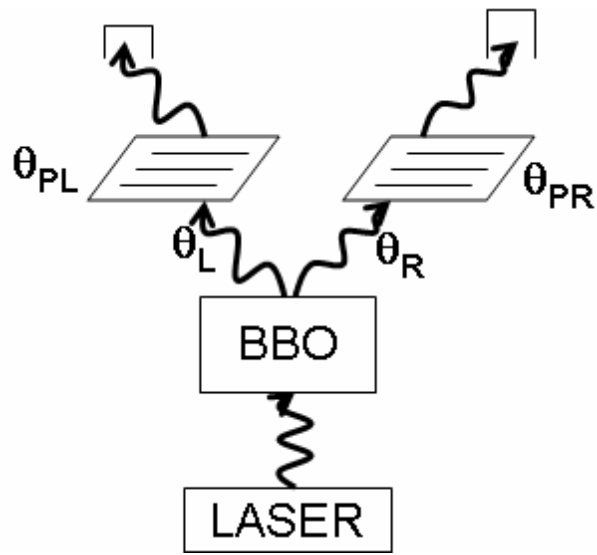

Figure 2. Schematic of a Simple "Bell's Theorem" Experiment

Let us consider the simple schematic in Figure 2. This schematic is not detailed enough to let us build one of these systems[3], but it is enough to represent the key issues of entanglement and nonlocality in the "Bell's Theorem" of Clauser et al[2].

The key to this experiment is the BBO crystal, where Type II Spontaneous Parametric Down Conversion takes place[27]. A laser is used to pump that crystal. With proper controls, the system will emit correlated pairs of linearly polarized photons into two channels – one photon to the left (L) and another to the right (R). More precisely, the pumped BBO acts as a source of two photons such that:

$$\theta_L = \theta_R + \tfrac{\pi}{2} \quad , \tag{10}$$

where $\theta_L$ is the polarization angle of the photon emitted to the left channel and $\theta_R$ is the polarization of the photon going to the right.

To analyze the Bell's Theorem effects, we need only consider five components in our "circuit": the BBO itself (equation 10), the two polarizers and the two counters on top. The counters are simply used to record (or, as seen in backwards time, emit) the "coincidence" of two photons coming out of the two channels "at the same time" ($t_+$). The purpose of the experiment is to record the rate at which we observe such two-photon arrivals at the counters at time $t_+$, $r(\theta_{PL},\theta_{PR})$, as a function of the polarization angles of the two polarizers. The weird outcome predicted by quantum mechanics and verified by experiment is that:

$$r(\theta_{PL},\theta_{PR}) = c * \cos^2(\theta_{PL} - (\theta_{PR} + \tfrac{\pi}{2})) \tag{11},$$

where c is a constant proportional to the rate of pair production in the BBO, held constant in the experiment.

The SPDC-2 systems developed by Yanhua Shih at the University of Maryland (with important help from Strekalov and Yoon-Ho Kim) provided more than just correlation of polarization. Their geometry was designed carefully to incorporate backwards-time concepts from Klyshko[28], who developed a concept of "biphoton." According to the biphoton concept, we may analyze a correlated pair of photons (as in figure 2) as a *single biphoton*, a photon which is emitted from the counter on one side backwards in time, which bounces off the BBO as from a mirror, and then proceeds to move forwards in time into the left-hand channel. Klyshko's concept allowed Shih to develop precise entanglement in momentum as well as in polarization; that in turn allowed him to perform precise experiments in quantum imaging which were far more precise at the time than anything else in the world.

What really exists across space-time in Figure 2, in cases where a coincidence is detected at the counters? In Klyshko's picture, it depends on which channel the biphoton is emitted from. If it is the right channel, there exists a photon of some polarization $\theta_R^+$ in the space-time zone between the right polarizer and the right counter; that is the polarization of light initially emitted from the right counter. There exists a photon of polarization $\theta_{PR}$ in the zone between the BBO and the right polarizer. There exists a photon of polarization $\theta_{PR}-(\pi/2)$ in the zone between the BBO and the left polarizer. And there exists a photon of polarization $\theta_{PL}$ in the zone between the left polarizer and the left counter.

This is not a satisfactory theory as it stands, because it says nothing about which channel would dominate. It actually offers two pictures, a right-channel picture and a left-channel picture, which imply different real states in different zones of the experiment! Nevertheless, it does reproduce equation 11, in a purely three-dimensional analysis.

To apply the MRF method numerically to this system, we need to know the constants $c_0$, $c_1$ and $c_2$ for both of the polarizers, and we need a representation of the two counters and the BBO. For the counters, it is enough to assume $Pr(\theta^+)=c$, some small uniform probability. For the BBO, it is enough to assume a uniform probability for $\theta_L$ or

$\theta_R$, and a *precise, rigid* enforcement of equation 10; in this case, it is the same joint probability distribution, no matter which channel we choose.

Even without doing a numerical calculation, it is easy to see what the picture is when $c_0 >> c_1, c_2$ and when the two polarizers have the same parameters (other than polarization angle of course). We predict a 50-50 probability distribution between two possible states within the experiment, in which one of the states is almost identical to the Klyshko right-channel picture and the other to his left-channel picture! The only difference is that the polarization of light between the "emitting counter" and the polarizer on the same channel will equal the angle of the polarizer, not a random angle, with very high probability. With a reasonable strong polarizer like a Polaroid film[18], we would expect $c_0$ to be something like three orders of magnitude larger than the other terms, enough to dominate the outcome; however, it may well be possible to measure the small changes in the traditional predictions implied by a more precise version of this kind of analysis.

# 3. THE BACKWARDS TIME TELEGRAPH (BTT)

Is it possible to build a communication device at some time t, which could be used to send a bit of information from time $t_+$ to time $t_-$, where $t_+ > t_- > t$? Unlike the orthodox time-forwards theories of physics, BTP does not immediately rule out this possibility[25]. I cannot yet say whether it is really possible or not, but a few comments may be still be in order.

In the first issue of *Physical Review Letters* in the year 2000, Yanhua Shih and collaborators described a quantum delay eraser experiment which almost appears o offer such a possibility. Under NSF funding, Shih successfully completed an experiment extending those effects over a much longer interval of time, using great lengths of optical fiber, similar in flavor to what is ongoing in Cramer's laboratory at the University of Washington. However, Shih's experiment could not actually give us a BTT, nor could systems based on Bell's Theorem, for reasons discussed by J.S. Bell[1].

Somewhat more interesting would be an extension of the "Popper's" experiment of Shih and Kim[29]. However, I have not had time to follow up on the possibilities there, except by posting some of the relevant discussions at www.werbos.com.

It is dangerous to rely too heavily on untrained intuition in situations like this. Yet – intuition suggests that a BTT would require us to find some sort of backwards-time free energy, a source of a kind of order which allows us to influence the past. If Price and Hawkings I are both right, a small but significant trickle of backwards-time free energy might be available from cosmological sources – perhaps even enough for a small bit of low-energy information processing. Or there might possibly be other approaches.

One device designer recently asked me: since electromagnetism is totally time-reversible, why can't be simply do a T reversal of systems that behave in forwards time the way we want to do things in backwards time? The problem is that the boundary conditions change when we do this (e.g. with regards to sources of free energy). Only a full-fledged MRF representation can specify the kinds of device properties we need to consider for successful reverse-time engineering.

# 4. CONCLUSIONS

This paper has argued that competing theories inspired by Bell's Theorem really are different theories of physics, with different and testable implications. They are not just a matter of interpretation and philosophy. Considerable work is needed to refine the various theories, and to design and perform the experiments needed to know which theories and which refinements really are correct. The fuzzy visionary consensus approach to physics, which tries to find the One Possible Super Theory of Physics through sheer deduction and papers over the differences, avoids the need to explore such tricky empirical issues – which is exactly why it is sterile. One way or another, it is time for a return to reality.

# APPENDIX: ENGINEERING THE DETECTION OF BACKWARDS TIME EFFECTS

This appendix will further address a question raised in section 2b, in regards to Huw Price's suggestion that we test the *original* version of Hawking's model of cosmology: i.e., how could we develop practical engineering methods to detect and perhaps even exploit "backwards time free energy"? How can we *find out* how difficult such methods would be to implement in experimental reality? I do not have a complete answer to this question as yet, but I can describe some of the possibilities.

    From the viewpoint of quantum physics, there is an analogy here to the old issue of "macroscopic Schrödinger cat." Superposition states were long known to be a fact of life down at the quantum level, but many top theorists once argued that it would be impossible to bring such effects to the macroscopic level, where they could be observed directly. It was very difficult to do this -- but, after an explicit understanding of the difficulties and the unknowns, Nakamura's group was able to develop a design to test quantum superposition at the macroscopic level, which proved that it is indeed possible[30]. In a similar manner, this appendix asks how we might develop designs that could test whether the time-symmetry which exists in quantum dynamics may be made observable at the macroscopic level.

    In engineering terms, the first challenge here is to develop photonic or electro-optic *circuits* which are capable of detecting when a source of energy (such as light received at night on the dark side of the moon) contains some amount of backwards-direction free energy, and not just the usual pure time-forwards free energy that we assume.

Again, an analogy may be necessary, to help us "think out of the box," and break out of the old "double standard[17]" in thinking about time. An analogy may be necessary, in helping us to properly establish boundary conditions and translate our new ideas about time into the required new models.

The challenge addressed here is analogous to the challenge of moving from Direct Current (DC) electronics to Alternating Current (AC) electronics. Here we are trying to move past the limitations of Forwards Time (FT) engineering to Alternating Time (AT) electronics.

In DC electronics, we could basically analyze any circuit by using the original form of Ohm's law. It was enough to know one number -- the resistance -- to characterize any linear circuit element. Nonlinear elements could be characterized by one real scalar function, I(V). The power source itself would be a DC source like a battery. To move on to AC electronics, it was necessary to go back to the entire menu of possible circuit elements, and develop a more complete and complex characterization. Even for linear circuit elements, we had to measure the resistance and reactance -- *two* numbers -- as they vary with frequency.

Here, in AT electronics, our first problem is with the power supply. In FT electronics, the socket we plug into provides either FT/AC or FT/DC power. But in AT electronics or electro-optics, we need to consider power sources (electronic or photonic) which may be a *mixture* of FT and backwards time (BT) power.

Huw Price suggests[17] that the optical energy from certain small directions of the sky might be more than ninety percent Backwards Time free energy, i.e. BT power. It may indeed be possible to go directly to the design of a highly focused, realistic telescope to detect FT free energy as such, from a very narrow window of the night sky. Here, however, I will begin by asking about larger sources of energy. For example, what happens if we build a passive device which receives a 50-50 *mixture* of FT and BT power from an external source? What if we built a device on the dark side of the moon to collect ALL of the light coming from a very large angle beyond our galaxy? How could we even detect the difference between such a mixed source of energy and the lack of any power at all?

A key concept here is the concept of a (temporally) *passive* device. In section 2d, in analyzing the polarizers in a "Bell's Theorem" experiment, I stressed that these circuit elements must be given a time-symmetric description, because their operation is not defined by use of a time-forwards reservoir of energy. Light from past or future may flow through a polarizer, but the polarizer does not have to be plugged into an FT/AC power source in order to do its job. There is not a hidden source of power, to make the polarizer work as it does.

When we receive a mixed source of FT and BT energy, the obvious engineering strategy is to try to *rectify* this energy somehow, into an FT stream and a BT stream. But I have not yet figured out how we might accomplish this. I have asked myself: what would happen if we were to plug in a simple rectifier into a "socket" which contains a 50-50 source of mixed FT and BT power?

It is easier for me to think about old-style diode power tubes here. Old style diode power tubes are basically passive devices. In an old style rectifier tube, electrons can only move in measurable numbers from the cathode to the anode, because the cathode is made of special materials and is hot (allowing an easy release of electrons) and the anode is not. But with a source of pure BT power, it would be equally easy for the electrons to be "emitted" in backwards time from the cathode; this would result in a current opposite to the usual direction, exactly canceling out the measurable effect, if the BT and FT power both happen to be DC.

More modern semiconductor rectifiers may or may not be completely different from old-style tube rectifiers, as AT circuit elements, even though they are equivalent in simple FT circuits. I simply do not yet know. Charge coupled detectors and avalanche devices are certainly not all passive devices. Thus I would predict that some "energy detection" devices would detect zero energy when plugged into a mixed FT/BT power supply, while others would detect nonzero energy.

For the case of a "BT" telescope (as in Price's suggestion), it may be possible to design a "time-inverted" avalanche detector, powered by FT free energy but highly responsive to pure BT photons. Or, following the Klyshko picture of Bell's Theorem experiments based on SPDC-II, we could use a pumped nonlinear crystal to "reflect" BT photons from a telescope into forwards time, where an ordinary avalanche detector could be used to detect their "reflection."

Going back to the case of a 50-50 FT/BT power supply -- what happens if we plug in a simple light bulb? IF a linear analysis were sufficient (which it may or may not be), then a light bulb plugged into a mixed 50-50 FT/BT power supply would emit ordinary photons moving outwards from the light bulk in forwards time, but would also emit photons in backwards time; the latter aspect would appear in forwards time like a "vacuum cleaner" pulling photons out from the environment.

If the bulb had a narrow filament -- as in a fuse -- we might indeed expect the fuse to burn out if the power supply is strong enough. This would be one way to detect the power supply. Does this violate time-symmetry? Not if we test the fuse to make sure that it works, and *then* plug it into the power supply. However, it is not obvious that a linear analysis is sufficient here; a more explicit model of forwards and backwards electron transport is called for. Still, if we find a power source which *appears* dark by some measures, and yet is capable of blowing out a fuse, that by itself would be *sufficient* to prove that we have located a mixed power source.

Of course, if we were to hook the power source up to a switch, the initial flows before and after we open the switch would be different for the FT and BT components. A meter away from the power source, the difference would be a matter of nanoseconds, but nanosecond delays can be detected[31]. This may be the easiest approach for now to detect such a source of mixed power.